\documentclass[aps,prd,onecolumn,floatfix,noshowpacs,tightenlines,noshowkeys,superscriptaddress,amsmath,amssymb,nofootinbib]{revtex4}
\usepackage{amssymb,amsbsy,epsfig,color,graphicx}
\usepackage{color}
\usepackage{[longtable}
\usepackage{array}
\usepackage{dcolumn}   
\usepackage{cellspace}
\usepackage{mathtools}
\usepackage{amstext}
\usepackage{amssymb}
\usepackage{stmaryrd}
\usepackage{stackrel}
\usepackage{graphicx}
\usepackage{esint}
\usepackage[utf8]{inputenc}
\usepackage{blindtext}
\usepackage{float}
\restylefloat{table}
\usepackage{booktabs}
\usepackage{enumitem} 
\usepackage{bbold}

\usepackage{etoolbox} 
\usepackage{lipsum} 
\usepackage[capitalize]{cleveref}
\usepackage{multirow}
\usepackage[caption=false]{subfig}
\renewcommand\[{\begin{equation}}
\renewcommand\]{\end{equation}}

\newcommand{\ba}{\begin{eqnarray}}
\newcommand{\ea}{\end{eqnarray}}

\makeatletter

\appto{\appendix}{%
\@ifstar{\def\theequation@prefix{A.}}%
{}%
}
\makeatother

\begin{document}

\title{Nonlocal generalization of Galilean theories and gravity}

\author{Luca Buoninfante}
\affiliation{Dipartimento di Fisica "E.R. Caianiello", Universit\`a di Salerno, I-84084 Fisciano (SA), Italy}
\affiliation{INFN - Sezione di Napoli, Gruppo collegato di Salerno, I-84084 Fisciano (SA), Italy}
\affiliation{Van Swinderen Institute, University of Groningen, 9747 AG, Groningen, The Netherlands}

\author{Gaetano Lambiase}
\affiliation{Dipartimento di Fisica "E.R. Caianiello", Universit\`a di Salerno, I-84084 Fisciano (SA), Italy}
\affiliation{INFN - Sezione di Napoli, Gruppo collegato di Salerno, I-84084 Fisciano (SA), Italy}

\author{Masahide Yamaguchi}
\affiliation{Department of Physics, Tokyo Institute of Technology, Tokyo 152-8551, Japan}


\begin{abstract}
In this paper we propose a wider class of symmetries including the Galilean shift symmetry as a subclass. We will show how to construct ghost-free nonlocal actions, consisting of infinite derivative operators, which are invariant under such symmetries, but whose functional form is not simply given by exponentials of entire functions. Motivated by this, we will consider the case of a scalar field and discuss the pole structure of the propagator which has infinitely many complex conjugate poles, but satisfies the tree-level unitarity. We will also consider the possibility to construct UV complete Galilean theories by showing how the ultraviolet behavior of loop integrals can be ameliorated. Moreover, we will consider kinetic operators respecting the same symmetries in the context of linearized gravity. In such a scenario, the graviton propagator turns out to be ghost-free and the spacetime metric generated by a point-like source is nonsingular. These new nonlocal models can be seen as an infinite derivative generalization of Lee-Wick theories and open a new branch of nonlocal theories.
\end{abstract}

\maketitle


\section{Introduction}

Back in the fifties, when the renormalizability of quantum electrodynamics was still not totally established, physicists were also trying to improve the ultraviolet (UV) behavior of the loop integrals by introducing nonlocal form factors in the Lagrangians \cite{Yukawa:1950eq}. It was observed that integrals could be made more convergent by using non-polynomial differential operators, and infinite order derivative\footnote{In this paper we will use both the expressions "nonlocal" and "infinite derivative" to mean the same thing.} actions with such form factors were also studied from a pure mathematical and axiomatic point of view \cite{efimov}. 

However, only in the last three decades infinite derivative models have aroused much interest and have been widely investigated \cite{Krasnikov,Kuzmin,Moffat,Tomboulis:1997gg,Biswas:2005qr,Modesto:2011kw,Biswas:2011ar,Biswas:2014yia,Tomboulis:2015gfa,Gama:2018cda,Hashi:2018kag,Buoninfante:2018mre}. Important results have been found especially in the context of string-field theory and $p$-adic string \cite{Witten:1985cc,Freund:1987kt,eliezer,Tseytlin:1995uq,Siegel:2003vt,Witten:2013pra,sen-epsilon}, and for quadratic (in the curvature) theories of gravity around Minkowski~\cite{Biswas:2005qr,Modesto:2011kw,Biswas:2011ar}, deSitter (dS) and anti-deSitter (AdS) backgrounds \cite{Biswas:2016etb}.

Infinite derivative gauge theories and their UV behavior were first studied in Refs.\cite{Krasnikov,Moffat,Tomboulis:1997gg}, where the authors computed the propagator and noticed that the presence of {\it exponentials of entire functions} does not introduce any new poles other than the local ones. Subsequently, the same results were also obtained in the context of gravity \cite{Biswas:2005qr,Modesto:2011kw,Biswas:2011ar}, where it is possible to formulate an infinite derivative theory whose action is of higher order in the curvature invariants but still free from ghost fields, therefore unitary at the quantum level \cite{sen-epsilon,carone,chin,Briscese:2018oyx}. Such theories are also often called {\it ghost-free infinite derivative field theories}.
An important point to note here is that the infinite order derivatives make the interaction nonlocal, indeed the interaction vertex is smeared out in the momentum space. 

In ghost-free infinite derivative gravity the propagator has been computed around  maximally symmetric backgrounds \cite{Biswas:2011ar,Biswas:2016etb}.  It was shown that no other dynamical degree of freedom is introduced besides the massless graviton of general relativity (GR) \cite{Krasnikov,Modesto:2011kw,Biswas:2011ar,Biswas:2016etb}. At the classical level, the nonlocal gravitational interaction can solve blackhole \cite{Modesto:2011kw,Biswas:2011ar,Biswas:2013cha,Biswas:2016etb,Edholm:2016hbt,Frolov,Frolov:2015bia,Frolov:2015usa,Buoninfante,Koshelev:2017bxd,Buoninfante:2018xiw,Koshelev:2018hpt,Buoninfante:2018rlq,Buoninfante:2018stt,Giacchini:2018wlf,Buoninfante:2018xif,Boos,Buoninfante:2019swn} and cosmological singularities \cite{Biswas:2005qr,Biswas:2010zk,Biswas:2012bp,Koshelev:2012qn,Koshelev:2018rau}. Infinite derivative gravity in $3$ dimensions has been recently studied, in both massless and massive cases; see Ref.\cite{Mazumdar:2018xjz}.  At a quantum level there are hints to show that the
UV behavior of the theory is ameliorated \cite{Kuzmin,Tomboulis:1997gg,Modesto:2011kw,Modesto:2014lga,Talaganis:2014ida,Biswas:2014yia}. Furthermore, in Ref.\cite{Ghoshal:2017egr} it was noticed that in an infinite derivative extension of the Abelian Higgs model has no instabilities emerge as the $\beta$-function vanishes in the UV regime, i.e. for energies beyond the scale of nonlocality, $p^2>M_s^2.$ Infinite derivative Lagrangians were also studied in the context of thermal field theory \cite{Biswas:2009nx,Biswas:2010xq,Biswas:2010yx}, inflationary cosmology \cite{inflation}, supersymmetry \cite{Giaccari:2016kzy,Kimura:2016irk} and applied to the study of the Casimir effect in curved background \cite{Buoninfante:2018bkc}. 
Recently, it has also been pointed out that in presence of multi particle interactions, such as mutiparticle scattering, the nonlocal scale can be transmuted from the UV to the IR depending on the number of particles involved in the scattering process. The emergence of such a new scale in the IR is an interesting 
result which demonstrates the existence of some  complementarity principle in infinite derivative theories~\cite{Buoninfante:2018gce}.

So far we have only mentioned nonlocal theories whose Lagrangians are constructed in terms of analytic operators. It is worthwhile to mention that nonanalytic operators like $1/\Box$ and ${\rm ln}(\Box)$ naturally emerge from one-loop quantum corrections to the effective action of quantum gravity \cite{Bravinsky,Deser:2007jk,Belgacem:2017cqo,Woodard:2018gfj}, and also in causal-set theory \cite{belenchia}. 

In this paper, our aim is to understand if there exist nonlocal actions which exhibit some special symmetries. In particular, we will ask whether we can enlarge the {\it Galilean symmetry} \cite{Nicolis:2008in}, $\partial_{\mu} \phi\longrightarrow\partial_{\mu} \phi+b_{\mu},$ which is typical of theories with some specific derivative self-interactions and is often taken as a fundamental guiding principle to construct and constrain the interaction Lagrangians in both flat \cite{Dvali:2000hr,Nicolis:2008in} and curved backgrounds \cite{Deffayet:2009wt,Deffayet:2010qz,Kobayashi:2010cm,Burrage:2010cu,Creminelli:2010qf,Deffayet:2011gz,Kobayashi:2011nu,Banerjee:2017qdl}. A very interesting aspect of higher derivative theories with Galilean symmetry is that the field equations are second order in the derivatives, thus no instabilities arise \cite{Nicolis:2008in}. However, we now wish to generalize to actions with derivatives of infinite order.

The paper is organized as follows. 
\begin{itemize}
	
	\item In Section \ref{enlarging} we will
	find nonlocal operators which are invariant under a larger
	class of transformations containing the Galilean transformations as a
	subclass.  
	
	\item In Section \ref{new-nonlocal}, we will introduce a new
	nonlocal field theoretical model motivated by such a new symmetry; we
	will compute the propagator and discuss the pole structure, by showing
	tree-level unitarity. We will show that the UV properties of $1$-loop integrals are
	ameliorated by the presence of derivatives of infinite order. Furthermore, we will notice that such nonlocal models can be seen as an infinite derivative generalization of the Lee-Wick theories, since the propagator possesses infinite complex conjugate poles.
	
	\item In Section \ref{new-gravity}, we will consider the same class of
	nonlocal operators in the context of linearized quadratic gravity, and also in this case we will compute the propagator by showing tree-level unitarity. Very interestingly, we will notice that it is possible to enlarge the class of nonlocal form factors which make the
	graviton propagator ghost-free at the tree-level, without restricting
	ourself to exponentials of entire functions. Moreover, we will show
	that the linearized spacetime metric generated by a point-like source
	is nonsingular. 
	
	\item In Section \ref{conclus}, we will discuss our results and draw the conclusions. 
	
	\item In Appendix \ref{append-shift}, we will discuss another larger class of transformations containing the shift
	transformations as a subclass, and consider a natural nonlocal extension of the Dirac action, which can be gauge invariant.
	
\end{itemize}
Throughout the paper we will work with the mostly positive metric signature $(-+++),$ and we use Natural Units $\hbar=1=c.$


\section{Enlarging the Galilean symmetry: from local to nonlocal operators}\label{enlarging}

In standard local field theory, Lagrangians are constructed in terms of polynomials of fields and polynomials of derivatives of fields since one is interested in observables at
low energies, therefore, the order of derivatives is always finite:
\begin{equation}
\mathcal{L}\equiv\mathcal{L}\left(\phi,\partial\phi,\partial^2\phi,\dots,\partial^n\phi\right),\label{local-lagr}
\end{equation}
where $n$ is a positive finite integer and $\phi(x),$ in principle, can be any kind of tensorial field. It is often very important to ask what kind of transformations leave a Lagrangian (action) invariant, especially in order to find symmetries and conserved quantities. In particular, we are interested here in the {\it Galilean shift symmetry} in the Minkowski spacetime, defined by the following transformation:
\begin{equation}
\partial_{\mu}\phi(x)\longrightarrow \partial_{\mu}\phi(x)+b_{\mu},\label{galilean-transf}
\end{equation}
which is also equivalent to
\begin{equation}
\phi(x)\longrightarrow \phi(x)+b_{\mu}x^{\mu}+a,\label{galilean-transf-2}
\end{equation}
where $\left\lbrace b_{\mu},a \right\rbrace $ are five constant parameters which generate the whole Galilean symmetry.
Lagrangians invariant under the above transformations can be constructed, in terms of a massless scalar field $\phi(x)$, where only specific derivative self-interactions are present. For example, the following Lagrangian exhibits such a symmetry \cite{Nicolis:2008in}:
\begin{equation}
\begin{array}{rl}
\mathcal{L}=&\displaystyle \frac{1}{2}\phi\Box\phi-\frac{1}{2\Lambda^3}\Box\phi\partial_{\mu}\phi\partial^{\mu}\phi+\cdots
\end{array}
\label{galilean-lagr}
\end{equation}
where $\Lambda$ can be, for example, a cut-off energy scale above which an effective field theory description may break down. A more general form of the Lagrangian in Eq.\eqref{galilean-lagr} can be found in Ref.\cite{Nicolis:2008in}. By integrating by parts it can be easily shown that the Lagrangian (or, more rigorously, the action) in Eq.\eqref{galilean-lagr} is invariant under the Galilean transformation in Eqs.(\ref{galilean-transf},\ref{galilean-transf-2}).

We now wish to understand whether we can enlarge the Galilean symmetry while working with nonlocal differential operators. First of all, for the sake of clarity, let us mention that a nonlocal Lagrangian is a function which can be also made up of non-polynomial differential operators, like for instance
\begin{equation}
\mathcal{L}\equiv\mathcal{L}\left(\phi,\partial\phi,\partial^2\phi,\dots,\partial^n\phi,\frac{1}{\Box}\phi,{\rm ln}\left( \Box/M_s^2\right)\phi,e^{\Box/M_s^2}\phi,\dots\right),\label{nonlocal-lagr}
\end{equation}
where the non-polynomial operators contain infinite order covariant derivatives; $M_s$ is the energy scale of nonlocality beyond which new physics should manifest and observables at high energy can be computed, and it is mathematically needed to make the arguments of logarithm and exponential dimensionless. For example, in the case of the exponential of the d'Alembertian, we can write the operator:
\begin{equation}
e^{\Box/M_s^2}=\sum\limits_{n=0}^{\infty}\frac{1}{n!}\left(\frac{\Box}{M_s^2}\right)^n,\label{taylor-exponential}
\end{equation}
where the derivative order $n$ goes up to infinity. As also mentioned above in the Introduction, exponentials of entire functions, like the one in Eq.\eqref{taylor-exponential}, have been used previously in the context of infinite derivative field theory and gravity \cite{Krasnikov,Kuzmin,Tomboulis:1997gg,Biswas:2005qr,Biswas:2011ar,Modesto:2011kw,Buoninfante:2018mre}. 

In order to satisfy  the Galilean shift symmetry in Eqs.(\ref{galilean-transf},\ref{galilean-transf-2}), we need to construct other nonlocal operators which are different from the ones already known in the literature. Further note that second order derivative operators, containing terms like the ones in Eq.\eqref{galilean-lagr}, are invariant under the Galilean transformation in Eqs.(\ref{galilean-transf},\ref{galilean-transf-2}). It is easy to understand that if $\Box \phi$ is invariant under the Galilean transformation, then any power $\Box^n \phi$ will be also invariant, which in turn implies that the following nonlocal operator is also invariant under the Galilean transformation:
\begin{equation}
\mathcal{O}_1\phi\equiv\left(e^{-\Box/M_s^2}-1\right)\phi=\sum\limits_{n=1}^{\infty}\frac{1}{n!}\left(-\frac{\Box}{M_s^2}\right)^n\phi,\label{nonlocal operator}
\end{equation}
which is slightly different from the one in Eq.\eqref{taylor-exponential}.

We now wish to ask the following question - is there any function $\psi(x)$, such that the transformation $\phi(x)\longrightarrow\phi(x)+\psi(x),$ leaves invariant the nonlocal operator in Eq.\eqref{nonlocal operator}, but {\it not} the local one? Or, in other words, can we find a function $\psi(x)$ such that $\left(e^{-\Box/M_s^2}-1\right)\psi=0,$ but $\Box\psi\neq 0$?

\subsection{1D case}

Let us start considering a $1$-dimensional case as a warm-up exercise, namely let us find solutions to the following nonlocal differential equation:
\begin{equation}
\left(e^{-\partial^2_x/M_s^2}-1\right)\psi(x)=0 \Longleftrightarrow e^{-\partial_x/M_s^2}\psi(x)=\psi(x), \label{nonlocal diff-1D}
\end{equation}
which also means to find the eigenfunctions $\psi$ of the operator $e^{-\partial_x^2/M_s^2}$ of unit eigenvalue.
First of all, note that the following property holds 
\begin{equation}
e^{-\partial^2_x/M_s^2}=e^{-\partial^2_x/M_s^2+i2\pi k}\equiv e^{\theta_{2}}, \label{property}
\end{equation}
where $k$ is an integer number and we have defined the differential operator
\begin{equation}
\theta_{2}:=-\frac{\partial^2_x}{M_s^2}+i2\pi k. \label{1-diff-oper}
\end{equation}
Thus, we need to find solutions for the equation $\theta_{2}\psi(x)=0:$
\begin{equation}
\theta_{2} \psi(x)=-\frac{\partial^2_x}{M_s^2}\psi(x)+i2\pi k\psi(x)=0\,\Longrightarrow \,\psi_k(x)=C_k e^{(1+i)\sqrt{\pi k} M_sx}+D_k e^{-(1+i)\sqrt{\pi k} M_sx}. \label{1-sol}
\end{equation}
where $C_k$ and $D_k$ are two integration constants, which need to be fixed by imposing that for $k=0$, then we recover the Galilean shift symmetry in Eqs.(\ref{galilean-transf},\ref{galilean-transf-2}). Indeed, by choosing
\begin{equation}
C_k=\frac{B}{\sqrt{k}},\,\,\,\,\,\,\,D_k=-\frac{B}{\sqrt{k}}+a,
\end{equation}
we obtain
\begin{equation}
\psi_k(x)=\frac{B}{\sqrt{k}}\left(e^{(1+i)\sqrt{\pi k} M_sx}- e^{-(1+i)\sqrt{\pi k} M_sx}\right)+ae^{-(1+i)\sqrt{\pi k} M_sx}, \label{boundary-cond}
\end{equation}
which in the limit $k\rightarrow 0$ gives $\psi_0(x)=bx+a:$
\begin{equation}
\lim\limits_{k\rightarrow0}\frac{B}{\sqrt{k}}\left(e^{(1+i)\sqrt{\pi k} M_sx}- e^{-(1+i)\sqrt{\pi k} M_sx}\right)=2(1+i)B\sqrt{\pi}M_sx\equiv bx,\label{boundary-cond-2}
\end{equation}
where $b\equiv2(1+i)B\sqrt{\pi}M_s.$  The solutions in Eq.\eqref{boundary-cond} are valid for positive integer, $k>0,$ while in the case of $k<0,$ the solutions read:
\begin{equation}
\psi_{k}(x)=-\frac{iB}{\sqrt{|k|}}\left(e^{(i-1)\sqrt{\pi |k|} M_sx}- e^{-(i-1)\sqrt{\pi |k|} M_sx}\right)+ae^{-(i-1)\sqrt{\pi |k|} M_sx}. \label{boundary-cond-k-negative}
\end{equation}
%

\subsection{4D case}

We can now apply the same procedure to the $4$-dimensional case. Also in this case we will find a larger class of solutions which are eigenfunctions of the differential operator $e^{-\Box/M_s^2}$ with eigenvalue equal to one,
\begin{equation}
\left(e^{-\Box/M_s^2}-1\right)\psi(x)=0 \Longleftrightarrow e^{-\Box/M_s^2}\psi(x)=\psi(x), \label{nonlocal diff-4D}
\end{equation}
and such that $\Box\psi(x)\neq 0.$ We can write
\begin{equation}
e^{-\Box/M_s^2}=e^{-\Box/M_s^2+i2\pi k}\equiv e^{\theta_{2}}, 
\end{equation}
where we have now defined
\begin{equation}
\theta_{2}:=-\frac{\Box}{M_s^2}+i2\pi k. \label{2-diff-oper}
\end{equation}
We can show that the solutions of the differential equation $\theta_{2} \psi(x)=0$ are given by:
\begin{equation}
\psi_k(x)=\frac{B}{\sqrt{k}}\left(e^{(1+i)\sqrt{\pi k} M_sc_{\mu}x^{\mu}}- e^{-(1+i)\sqrt{\pi k} M_sc_{\mu}x^{\mu}}\right)+ae^{-(1+i)\sqrt{\pi k} M_sc_{\mu}x^{\mu}}, \label{boundary-cond-4D}
\end{equation}
where $c^2\equiv c_{\mu}c^{\mu}=1;$ the analog solution for $k<0$ can be easily found too, as done above for the $1$-dimensional case.
In the limit $k\rightarrow 0$ we obtain:
\begin{equation}
\lim\limits_{k\rightarrow0}\psi_k(x)=2(1+i)B\sqrt{\pi}M_sc_{\mu}x^{\mu}+a\equiv b_{\mu}x^{\mu}+a,\label{limit-4D}
\end{equation}
which recovers the Galilean shift symmetry in Eqs.\eqref{galilean-transf-2}, with $b_{\mu}\equiv 2(1+i)B\sqrt{\pi}M_sc_{\mu},$ where the integration constant can be also written as $B=\sqrt{b_{\mu}b^{\mu}}/(2(1+i)\sqrt{\pi}M_s).$ Note that we can now rewrite the function $\psi_{k}$ in Eq.\eqref{boundary-cond-4D} in a more compact form as follows:
\begin{equation}
\psi_k(x)=\displaystyle \frac{2B}{\sqrt{k}}{\rm sinh}\left(\frac{\sqrt{k}}{2B}b_{\mu}x^{\mu}\right)+ae^{-\frac{\sqrt{k}}{2B}b_{\mu}x^{\mu}}. \label{compact-form}
\end{equation}
Hence, we have explicitly shown that by working with the nonlocal, infinite derivative operator introduced in Eq.\eqref{nonlocal operator}, we can enlarge the Galilean shift symmetry, which now becomes a subclass of a wider class described by the following family of parameters: 
\begin{equation}
\left\lbrace a,b_{\mu},k\right\rbrace .\label{family-parameters}
\end{equation}
Thus, the Galilean transformations correspond to the subfamily $\left\lbrace a,b_{\mu},0\right\rbrace.$ In other words, the Galilean shift symmetry turns out to be a subclass $(k=0)$ of this larger symmetry which is expressed in terms of the following field transformation\footnote{The transformation in Eq.\eqref{larger-symmetry} generalizes the Galilean symmetry in Eq.\eqref{galilean-transf-2} to the case of non-polynomial differential operators. In a similar way, we can also construct other nonlocal differential operators which exhibit an enlarged shift symmetry ($\phi\longrightarrow \phi+c$). See Appendix \ref{append-shift} for more details.}:
\begin{equation}
\phi(x)\longrightarrow \phi(x)+\psi_k(x).\label{larger-symmetry}
\end{equation}
%

\subsection{Generic powers of $\Box$}

So far we have only performed our study for the nonlocal operator $\mathcal{O}_1=\left(e^{-\Box/M_s^2}-1\right),$ where the exponent is simply given by $\Box.$ However, we can also consider more generic entire functions in the exponent, as for example $(-\Box/M_s^2)^n$ or even non-polynomial entire functions. In the former general scenario, a new symmetry can be still found, but the solutions for $\psi_k(x)$ become more complicated as the power $n$ increases.

For instance, we can consider the following more general operator:
\begin{equation}
\mathcal{O}_n:=e^{\left(-\Box/M_s^2\right)^n}-1. 
\end{equation}
First of all, we can write
\begin{equation}
e^{\left(-\Box/M_s^2\right)^n}=e^{\left(-\Box/M_s^2\right)^n+i2\pi k}\equiv e^{\theta_{n}}, 
\end{equation}
where 
\begin{equation}
\theta_{n}:=\left(-\frac{\Box}{M_s^2}\right)^n+i2\pi k.
\end{equation}
In this more general case the solutions of the field equation 
\begin{equation}
\theta_{n}\psi_k(x)\equiv \left(-\frac{\Box}{M_s^2}\right)^n\psi_k(x)+i2\pi k\psi_k(x)=0\label{diff-eq-n}
\end{equation}
is formally given by:
\begin{equation}
\psi_k(x)=\sum\limits_{l=1}^{2n}C_{k,l}e^{d_l(k)^{1/2n}M_sc_{\mu}x^{\mu}},
\end{equation}
where $d_l$ are constant parameters depending on the order $2n$ of the differential equation in Eq.\eqref{diff-eq-n} and satisfy the algebraic equation $(-d_l^2)^n+2\pi i=0$, while $C_{k,l}$ are integration constants. Anyway, in this paper we will mainly work with the nonlocal operator $\mathcal{O}_1.$


\section{A nonlocal model with infinite complex conjugate poles}\label{new-nonlocal}

In this Section we wish to construct actions, or in other words Lagrangians, which respect the new symmetry in Eq.\eqref{larger-symmetry}. By working with the nonlocal operator $\mathcal{O}_1,$ defined in Eq.\eqref{nonlocal operator}, possible action terms are given by
\begin{equation}
S=S_1+S_2+S_3+\cdots+ S_n+\cdots\label{action Sn}
\end{equation}
where 
\begin{equation}
\begin{array}{rl}
S_1=& \displaystyle -\frac{M_s^2}{2}\int d^4x\phi\left(e^{-\Box/M_s^2}-1\right)\phi,\\[2.5mm]
S_2=& \displaystyle -\lambda_2\frac{M_s^2}{2}\int d^4x\left[\left(e^{-\Box/M_s^2}-1\right)\phi\right]^2,\\[2.5mm]
S_3=& \displaystyle -\lambda_3\frac{M_s}{3!}\int d^4x\left[\left(e^{\Box/M_s^2}-1\right)\phi\right]^3,\\[2.5mm]
\vdots&\,\,\,\,\,\,\,\,\,\,\,\,\,\,\,\,\,\,\,\,\,\,\,\,\,\,\,\,\,\,\,\,\,\,\,\,\,\,\vdots\\
S_n=& \displaystyle -\lambda_n\frac{M_s^{4-n}}{n!}\int d^4x \left[\left(e^{\Box/M_s^2}-1\right)\phi\right]^n\\
\vdots&\,\,\,\,\,\,\,\,\,\,\,\,\,\,\,\,\,\,\,\,\,\,\,\,\,\,\,\,\,\,\,\,\,\,\,\,\,\,\vdots
\end{array}\label{action-terms Sn}
\end{equation}
with $\lambda_i$ being dimensionless coupling constants and the iteration is meant for $n>2$. For instance, up to cubic vertex interaction, the action in Eq.\eqref{action Sn} explicitly reads:
\begin{equation}
\begin{array}{rl}
S=& \displaystyle -\frac{1}{2}\int d^4x\left[(1-2\lambda_2)M_s^2\phi e^{-\Box/M_s^2}\phi-(1-\lambda_2)M_s^2\phi^2+\lambda_2M_s^2 \left(e^{-\Box/M_s^2}\phi\right)^2\right.\\[2.5mm]
&\displaystyle  \,\,\,\,\,\,\,\,\,\left.-\frac{\lambda_3M_s}{3}\phi^3 +\lambda_3M_s\phi^2e^{\Box/M_s^2}\phi-\lambda_3 M_s\phi\left(e^{\Box/M_s^2}\phi\right)^2+\frac{\lambda_3M_s}{3}\left(e^{\Box/M_s^2}\phi\right)^3 \right]+\cdots,
\end{array}\label{up-to-cubic}
\end{equation}
and it is invariant under the transformation in Eq.\eqref{larger-symmetry}. Moreover, we can expand the previous action in powers of $1/M_s,$ and up to order $\mathcal{O}(1/M_s^5)$ we obtain 
\begin{equation}
S=\int d^4x\left\lbrace \frac{1}{2}\phi\Box\phi-\frac{(1+2\lambda_2)}{4M_s^2}\phi\Box^2\phi+\frac{(1+6\lambda_2)}{12M_s^4}\phi\Box^3\phi-\frac{\lambda_3}{3!M_s^5}\left(\Box \phi\right)^3+\mathcal{O}\left(\frac{1}{M_s^6}\right)\right\rbrace,\label{action Sn-local-limit}
\end{equation}
which is now invariant under the Galilean shift transformation in Eqs.(\ref{galilean-transf},~\ref{galilean-transf-2}).
However, the nonlocal, infinite derivative  action in Eqs.(\ref{action Sn},~\ref{action-terms Sn}) is not the most general one. Indeed, we can also consider nonlocal terms which include the ones in Eq.\eqref{galilean-lagr} when expanding in powers of $1/M_s$. For example, the following nonlocal term
\begin{equation}
M_s^3\int d^4x \left(e^{\Box/M_s^2}-1\right)\phi\frac{\left(e^{\Box/M_s^2}-1\right)}{\Box}\partial_{\mu}\phi\frac{\left(e^{\Box/M_s^2}-1\right)}{\Box}\partial^{\mu}\phi,
\end{equation}
is invariant under the transformation in Eq.\eqref{larger-symmetry} and, the first nonvanishing term in the expansion is given by  
\begin{equation}
\frac{1}{M_s^3}\Box\phi \partial_{\mu}\phi\partial^{\mu}\phi,\label{locoal-inter-term}
\end{equation}
which appears in the effective local Lagrangian in Eq.\eqref{galilean-lagr}, where the energy cut-off is now the nonlocal scale $M_s.$
Hence, we have shown that by demanding the symmetry in Eq.\eqref{larger-symmetry} we can straightforwardly construct {\it new} field theoretical models for nonlocal interaction.


\subsection{Nonlocal propagator}

In this subsection, we will consider the simplest action possessing the symmetry in Eq.\eqref{larger-symmetry} and only work with a kinetic term, as we are mainly interested in computing the propagator. Such a kinetic action is
\begin{equation}
S_1= -\frac{M_s^2}{2}\int d^4x\phi\left(e^{-\Box/M_s^2}-1\right)\phi,
\label{kineti term}
\end{equation}
which, in the local limit, recovers the Klein-Gordon action for a massless scalar field. The field equation can be easily found by variating the action and reads:
\begin{equation}
e^{-\Box/M_s^2}\phi=\phi.
\label{field-eq}
\end{equation}
The bare propagator is defined as the inverse of the kinetic operator, and in momentum space it is given by
\begin{equation}
\Pi(p)=\frac{1}{M_s^2\left(e^{p^2/M_s^2}-1\right)}.
\label{propagator}
\end{equation}
First of all, because of the presence of infinite order time derivatives {\it no} spectral representation can be defined for the propagator in Eq.\eqref{propagator} since the time-ordered structure is lost. As a physical consequence, causality is violated at very small length scales of the order of $1/M_s\,;$ see for instance Refs.\cite{Tomboulis:2015gfa,Buoninfante:2018mre}. 

Note that the pole structure of the propagator is more complicated than the local one; besides the usual real massless pole at $p^2=0,$ we now have an {\it infinite number of complex conjugate poles}. Indeed, the denominator in Eq.\eqref{propagator} vanishes for 
\begin{equation}
p^2=i2\pi M_s^2 \ell \, \Longleftrightarrow\, p^0=\pm \sqrt{\vec{p}^2-i2\pi M_s^2 \ell},\label{poles}
\end{equation}
where $\ell$ is an integer number: $\ell=0$ corresponds to the only real massless pole, while each value of $\ell\neq 0$ is associated with two complex poles, whose conjugates are the ones corresponding to the opposite integer $-\ell.$ In fact, the square root in Eq.\eqref{poles} can be also decomposed in real and imaginary parts as follows:
\begin{equation}
p^0=\pm \left(\sqrt{\frac{\vec{p}^2+\sqrt{\vec{p}^4+4\pi^2M_s^4\ell^2}}{2}}-i\varepsilon(\ell)\sqrt{\frac{-\vec{p}^2+\sqrt{\vec{p}^4+4\pi^2M_s^4\ell^2}}{2}}\right),\label{poles-square-root}
\end{equation}
where $\varepsilon(\ell):=\theta(\ell)-\theta(-\ell),$ with $\theta$ being the Heaviside step function, or equivalently $\varepsilon(\ell)\equiv{\rm sign}(\ell).$ See Fig. $1$ for the graphic location of the poles. 
\begin{figure}[t]
	\includegraphics[scale=0.48]{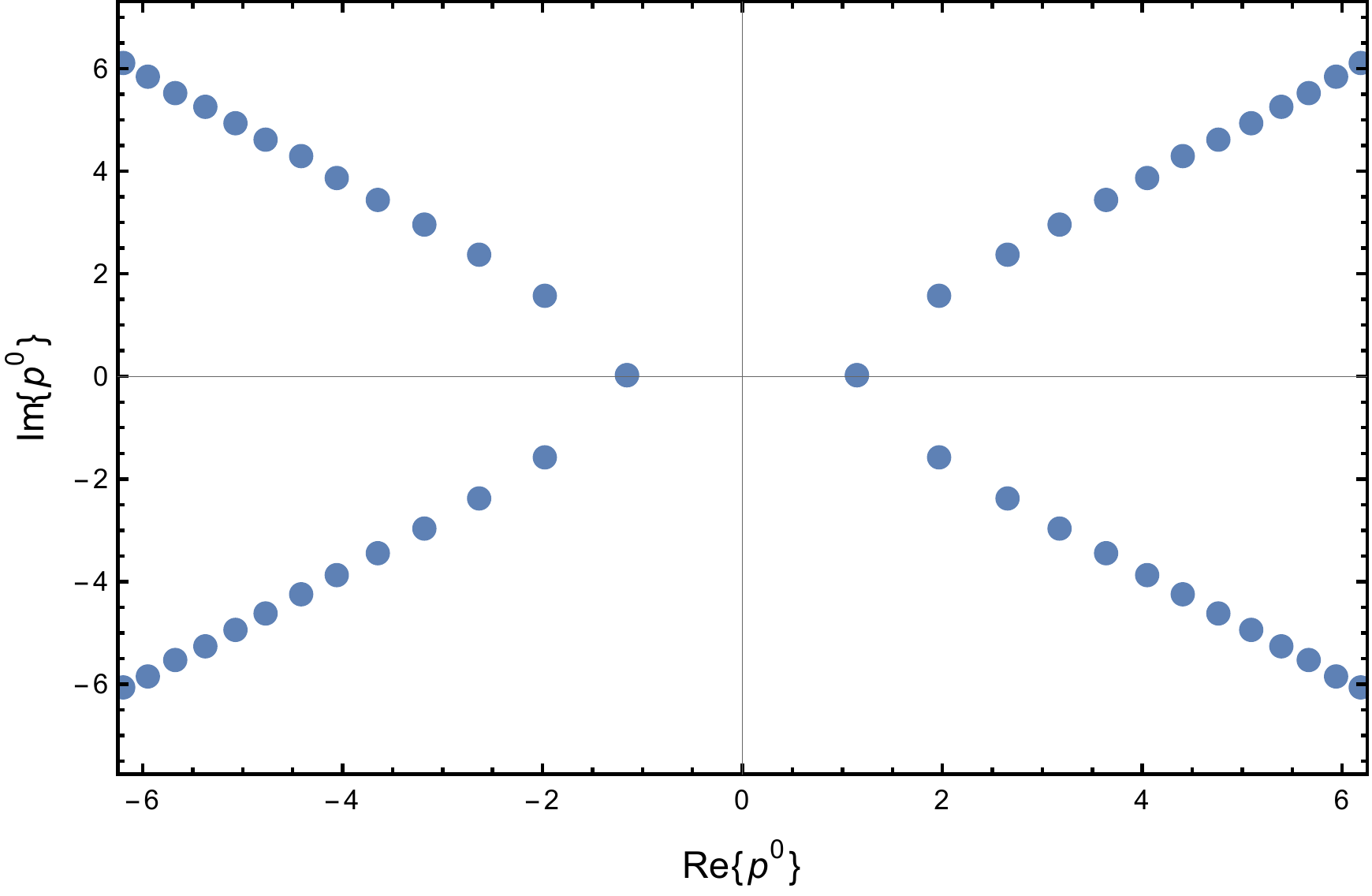}
	\centering
	\protect\caption{In this plot it is illustrated the complex plane for $p^0$ and the locations of some poles of the propagator in Eq.\eqref{propagator}. We have set $M_s=|\vec{p}|=1$ for simplicity, and plotted the poles $\ell=0,\dots,12.$}
\end{figure}

We can isolate the real massless pole in the propagator by writing:
\begin{equation}
\Pi(p)=\left(\frac{p^2}{M_s^2\left(e^{p^2/M_s^2}-1\right)}\right)\frac{1}{p^2}\equiv \frac{1}{f(p)}\frac{1}{p^2},\,\,\,\,\,\,\,\,\,\,f(p)\equiv M_s^2\frac{e^{p^2/M_s^2}-1}{p^2},
\label{propagator-2}
\end{equation}
where the zeros of the function $f(p)$ correspond to the complex poles. Indeed $f(p)=0$, if and only if $p^2=i2\pi M_s^2 \ell,$ with $\ell\neq 0.$ Furthermore, the propagator has {\it no essential singularities} at infinity which, in principle allow us to define a Wick rotation from Minkowski to Euclidean space and vice-versa.

Note that the presence of complex poles, in general, may spoil perturbative unitarity \cite{Anselmi:2016fid}, in such a way that predictability would be lost at the quantum level. For this reason, further investigations are needed in order to understand whether the unitarity condition, and so the optical theorem, are satisfied.


\subsection{Unitarity with infinite complex conjugate poles}\label{unitarity-subsection}

The unitarity condition of the $S$-matrix reads:
\begin{equation}
S^{\dagger}S=\mathbb{1},
\label{unitarity-cond}
\end{equation}
which, by writing $S=\mathbb{1}+iT,$ can also be expressed as follows
\begin{equation}
2{\rm Im}\left\lbrace T \right\rbrace=T^{\dagger}T;
\label{optical theorem}
\end{equation}
where the last equation represents the so called optical theorem. One of the implication of the optical theorem is that the imaginary part of any amplitude $T$ can not be negative. For instance, in the case of a tree-level amplitude with {\it constant} three-vertex interaction, it requires the imaginary part of the propagator to satisfy the inequality: ${\rm Im}\left\lbrace \Pi(p)\right\rbrace \geq 0.$
It is very important to understand whether the imaginary part of the propagator in Eq.\eqref{propagator} satisfies the optical theorem, i.e. whether the presence of infinite complex conjugate poles spoils unitarity or not.

Let us introduce a new variable $z:=-p^2/M_s^2$, so that the nonlocal propagator in Eq.\eqref{propagator} will read
\begin{equation}
\Pi(z)=\frac{1}{M^2_s\left(e^{-z}-1\right)},
\label{propag-z}
\end{equation}
whose poles are given by $z=i2\pi \ell,$ where $\ell$ is an integer number. We now wish to partially decompose the propagator in an infinite number of fractions, each corresponding to a single pole. Indeed, we can write
\begin{equation}
\begin{array}{rl}
\Pi(z)=\displaystyle -\frac{e^{z/2}}{M^2_s\left(e^{z/2}-e^{-z/2}\right)}
= \displaystyle-\frac{1}{2M^2_s}\frac{e^{z/2}}{{\rm sinh}(z/2)}
= \displaystyle\frac{1}{i2M_s^2}\frac{e^{z/2}}{{\rm sin}(iz/2)},
\end{array}
\label{parial-decomp}
\end{equation}
then by using the following identity \cite{SZILI}
\begin{equation}
\frac{1}{{\rm sin}(iz/2)}=\frac{2}{i}\sum\limits_{\ell=-\infty}^{\infty}(-1)^{\ell}\frac{1}{z+i2\pi \ell},
\label{identity}
\end{equation}
we obtain\footnote{Note that the result in Eq.\eqref{step-2} is a consequence of the Weierstrass factorization theorem for entire functions. Indeed, the theorem states that any entire function $H(z)$ can be always written as $H(z)=e^{h(z)}\prod\limits_{i=1}^{N}(z-\eta^2_i),$ where $h(z)$ is also an entire function and the number $N$ can be either finite or infinite, and counts the number of zeros of the function $H(z).$ In our case, $H(z)=2M_s^2e^{-z/2}{\rm sinh}(z/2)$ is the kinetic operator which has an infinite number of zeros given by $\eta^2_{\ell}=i2\pi M^2_s\ell,$ with $\ell\in \mathbb{Z}.$}
\begin{equation}
\Pi(z)=-\frac{e^{z/2}}{M^2_s z}-\frac{e^{z/2}}{M_s^2}\sum\limits_{\ell=1}^{\infty}(-1)^{\ell}\left(\frac{1}{z+i2\pi \ell}+\frac{1}{z-i2\pi \ell}\right).
\label{step-2}
\end{equation}
By going back to the momentum variable $p^2,$ we obtain the nonlocal propagator in Eq.\eqref{propagator} expressed in partial decomposition:
\begin{equation}
\Pi(p)=\frac{e^{-p^2/2M_s^2}}{p^2}+e^{-p^2/2M_s^2}\sum\limits_{\ell=1}^{\infty}(-1)^{\ell}\left(\frac{1}{p^2+i2\pi M_s^2 \ell}+\frac{1}{p^2-i2\pi M_s^2 \ell}\right).
\label{propag-partial}
\end{equation}
Therefore, we have been able to explicitly isolate all the poles of the propagators. The first term corresponds to the real massless pole, $p^2=0,$ while the second term takes into account of all the infinite complex conjugate poles, $p^2=\pm i2\pi M_s^2 \ell$ with $\ell\neq 0.$ By using this form of the propagator it is now easy to compute its imaginary part:
\begin{equation}
\begin{array}{rl}
{\rm Im}\left\lbrace \Pi(p)\right\rbrace =\displaystyle {\rm Im}\left\lbrace\frac{e^{-p^2/2M_s^2}}{p^2-i\epsilon}\right\rbrace
= \frac{e^{-p^2/2M_s^2}\epsilon}{p^4+\epsilon^2}=\pi \delta^{(4)}(p^2)>0,
\end{array}
\label{imaginary-part}
\end{equation}
where in the last step we have taken the limit $\epsilon \rightarrow 0$, and  we have used the fact that
\begin{equation}
{\rm Im}\left\lbrace \sum\limits_{\ell=1}^{\infty}(-1)^{\ell}\left(\frac{1}{p^2+i2\pi M_s^2 \ell}+\frac{1}{p^2-i2\pi M_s^2 \ell}\right)\right\rbrace =0.
\end{equation}
Thus, the infinite complex poles do not contribute to the imaginary part of the propagator, and it is due to the fact that they appear in {\it conjugate pairs}. This implies that the propagator in Eqs.(\ref{propagator},\ref{propag-partial}) satisfies the tree-level unitarity condition.

It is worthwhile to mention that in Refs.\cite{Lee:1969fy,Anselmi:2017ygm,Anselmi:2017yux,Anselmi:2018kgz,Modesto:2015ozb}, the authors have considered the unitarity issue with a finite number of complex conjugate poles; in particular in Refs.\cite{Anselmi:2017yux,Anselmi:2018kgz} it was rigorously shown that the optical theorem is preserved at all orders in pertubation theory. They have considered the so called Lee-Wick theories, whose Lagrangians are sixth order in the derivatives and the propagator is made up of the usual real pole plus a pair of complex conjugate poles. Our propagator in Eqs.(\ref{propagator},\ref{propag-partial}) can be seen as a nonlocal extension of the Lee-Wick propagator, where we now have infinite pairs of complex conjugate poles whose {\it infinite} number is related to the presence of {\it infinite} order derivatives in the Lagrangian. In this respect, the new nonlocal model we have introduced can be seen as an {\it infinite derivative Lee-Wick model}.

Since the nonlocal propagators whose form given by the first term in Eq.\eqref{propag-partial} leads to unitary theories \cite{sen-epsilon,carone,chin,Briscese:2018oyx}, and knowing that the presence of a finite number of complex conjugate poles does not break the unitary condition either, we would expect also in our case the optical theorem to hold at all orders in perturbation theory. However, further investigations on perturbative unitarity with infinite complex conjugate poles will be subject of future works.


\subsection{Ultraviolet behavior}

The tree-level Lagrangians in local Galilean theories are free from any kind of instabilities as the corresponding field equations turn out to be of second order in the derivatives \cite{Nicolis:2008in}. However, as shown in Refs.\cite{dePaulaNetto:2012hm,Brouzakis:2013lla}, quantum corrections can introduce new higher derivative terms like $\phi\Box^2\phi,$ $\phi\Box^3\phi$ and $\phi\Box^4\phi,$ which cause classical instabilities and unitarity violation, since ghost modes are introduced.

Nonlocal generalizations of Galilean theories can avoid any ghost degrees of freedom, not only at the tree-level but also when quantum loop corrections are taken into account, due to their improved UV behavior. As an example, we can consider the following nonlocal model\footnote{Note that the action in Eq. \eqref{action-general-nonlocal} is intrinsically nonlocal, i.e. there exist no field redefinition which can transform it into a local form.}:
\begin{equation}
\begin{array}{rl}
S=&\displaystyle -\frac{M^2_s}{2}\int d^4x \phi\left(e^{-\Box/M_s^2}-1\right)\phi\\
&\displaystyle  -\lambda\frac{M_s^3}{2}\int d^4 x \left(e^{\Box/M_s^2}-1\right)\phi \frac{\left(e^{\Box/M_s^2}-1\right)}{\Box}\partial_{\mu}\phi \frac{\left(e^{\Box/M_s^2}-1\right)}{\Box}\partial^{\mu}\phi,
\end{array}\label{action-general-nonlocal}
\end{equation}
where $\lambda$ is a dimensionless coupling constant. Very interestingly, in the low energy regime, $\Box\ll M_s^2,$ the action in Eq.\eqref{action-general-nonlocal} reduces to some version of local Galilean actions with the cubic term in Eq.\eqref{locoal-inter-term} plus higher order terms including the $1$-loop quantum corrections found in Refs.\cite{dePaulaNetto:2012hm,Brouzakis:2013lla}; for instance, by expanding up to $\mathcal{O}(1/M_s^6)$ we obtain:
\begin{equation}
\begin{array}{rl}
S=&\displaystyle \int d^4x\left\lbrace  \frac{1}{2}\phi\Box\phi-\frac{1}{4M_s^2}\phi\Box^2\phi-\frac{\lambda}{2M_s^3}\Box\phi\partial_{\mu}\phi\partial^{\mu}\phi\right.\\
&\displaystyle \,\,\,\,\,\,\,\,\,\,\,\,\,\,\,\,\,\,\, \left.+\frac{1}{12M_s^4}\phi\Box^3\phi- \frac{\lambda}{M_s^5}\Box\phi\partial_{\mu}\phi\Box\partial^{\mu}\phi-\frac{1}{24M_s^6}\phi\Box^4\phi+\mathcal{O}(1/M_s^7)\right\rbrace.
\end{array}\label{action-general-nonlocal-expansion}
\end{equation}
Therefore, the action in Eq.\eqref{action-general-nonlocal} can represent a UV completion of the local cubic Galilean action considered in Refs.\cite{dePaulaNetto:2012hm,Brouzakis:2013lla}; we will now clarify it with a $1$-loop computation.

The corresponding Feynman rules in Euclidean space are given by the nonlocal propagator $\Pi(k)$ in Eq.\eqref{propagator} and by the following three-vertex:
\begin{equation}
\begin{array}{rl}
V(k_1,k_2,k_3)=& \displaystyle \lambda M_s^3 \left(e^{-k_1^2/M_s^2}-1\right)\left(e^{-k_2^2/M_s^2}-1\right)\left(e^{-k_3^2/M_s^2}-1\right)\\
& \displaystyle  \times \left(\frac{k_1\cdot k_2}{k_1^2 k_2^2}+\frac{k_1\cdot k_3}{k_1^2 k_3^2}+\frac{k_2\cdot k_3}{k_2^2 k_3^2}\right).
\end{array}\label{3vertex}
\end{equation}
The presence of the exponentials in both propagator and interaction vertex ameliorates the UV behavior of the theory not only at the tree-level but also at higher loop orders. From power counting arguments we can easily understand how loop integrals will behave: since both propagator and vertices are exponentially suppressed at high energy, the UV behavior of the loop integrals is generally governed by $e^{-Ik^2/M_s^2},$ where $I$ is the number of internal propagators, thus the superficial degree of divergence is given by
\begin{equation}
D=-I.\label{degree-divergence}
\end{equation}
Note that $D$ is always negative, which is a good hint in favor of the finiteness of loop integrals for the model in Eq.\eqref{action-general-nonlocal}. We can explicitly show this feature, and make it more clear, by computing the self-energy $\Sigma(p)$ at $1$-loop for the action in Eq.\eqref{action-general-nonlocal}:
\begin{equation}
\begin{array}{rl}
\Sigma^{(1)}(p)=&\displaystyle \int \frac{d^4k}{(2\pi)^4} \Pi(k)\Pi(p-k)V^2(k,p-k,p)\\
= &\displaystyle  \lambda^2M_s^2\left(e^{-p^2/M_s^2}-1\right)^2\int \frac{d^4k}{(2\pi)^4} \frac{\left(e^{-k^2/M_s^2}-1\right)^2 \left(e^{-(p-k)^2/M_s^2}-1\right)^2}{\left(e^{k^2/M_s^2}-1\right) \left(e^{(p-k)^2/M_s^2}-1\right)} \\
& \,\,\,\,\,\,\,\,\,\,\,\,\,\,\,\,\,\,\,\,\,\,\,\,\,\,\,\,\displaystyle \times \left(\frac{k\cdot(p-k)}{k^2(p-k)^2}+\frac{k\cdot p}{k^2p^2}+\frac{p\cdot(p-k)}{p^2(p-k)^2}\right)^2.
\end{array}\label{self-energy}
\end{equation}
The integral for the self-energy is finite, indeed by taking the UV regime $(k^2\gg M_s^2)$ of the integrand we obtain
\begin{equation}
\displaystyle \Sigma^{(1)}(p)\xrightarrow{UV}\displaystyle \lambda^2M_s^2 \left(e^{-p^2/M_s^2}-1\right)^2 e^{-p^2/M_s^2} \int \frac{d^4k}{(2\pi)^4} \frac{e^{-2k^2/M_s^2}}{k^2},
\label{self-energy-UV}
\end{equation}
where the factor $2$ in the exponent is in agreement with the power counting in Eq.\eqref{degree-divergence}: $D=-2.$ Note that the integrals in Eq.\eqref{self-energy} can be computed in Euclidean space and then has to be analytically continued back to Minkowski.

From Eq.\eqref{self-energy-UV} we can notice that the loop integral is exponentially suppressed at high energy and does not need to be renormalized. Therefore, for the model in Eq.\eqref{action-general-nonlocal} we would expect no ghost degrees of freedom to emerge when quantum loop corrections are taken into account, unlike the local case studied in Refs.\cite{dePaulaNetto:2012hm,Brouzakis:2013lla}. However, more general studies regarding the UV behavior of these nonlocal models are needed and will be subjected to future investigations.


\subsection{Comparison with infinite derivative field theory (IDT)}

In the context of IDT, the simplest action we can consider is \cite{Tomboulis:2015gfa,Buoninfante:2018mre}
\begin{equation}
S_{\rm IDT}=\int d^4x \left(\frac{1}{2}\phi e^{-\Box/M_s^2}\Box\phi-\frac{\lambda}{3!}\phi^3\right),\label{IDT-action-1}
\end{equation}
which, by making the field redefinition $\tilde{\phi}=e^{-\Box/2M_s^2}\phi,$ can be equivalently written as 
\begin{equation}
S_{\rm IDT}=\int d^4x \left(\frac{1}{2}\tilde{\phi}\Box\tilde{\phi}-\frac{\lambda}{3!}\left(e^{\Box/2M_s^2}\tilde{\phi}\right)^3\right).\label{IDT-action-2}
\end{equation}
From Eqs.(\ref{IDT-action-1},\ref{IDT-action-2}) it is clear that nonlocality in IDT becomes important only when the interaction is switched on, while at the level of free theory it does not play any role and the only solutions are given by the local ones \cite{Barnaby:2007ve}, i.e. $\Box\phi=0.$ A different scenario happens in our nonlocal model, indeed the free field equation in Eq.\eqref{field-eq} admit a larger class of solution as we have shown above. 

The propagator for the action in Eq.\eqref{IDT-action-1} is given by:
\begin{equation}
\Pi_{\rm IDT}(p)=\frac{e^{-p^2/M_s^2}}{p^2}\equiv \frac{1}{f_{\rm IDT}(p)}\frac{1}{p^2},\,\,\,\,\,\,\,\,\,\,f_{\rm IDT}(p)\equiv e^{p^2/M_s^2}\label{propag-stringy}
\end{equation}
which has only one massless real pole at $p^2=0,$ and has the same form of the first term of the nonlocal propagator in Eq.\eqref{propagator}. Therefore, the main difference between the propagators in Eq.\eqref{propag-stringy} and Eq.\eqref{propagator} is due to the presence of infinite complex conjugate poles in the latter, corresponding to the zeros of the function $f(p)$ in Eq.\eqref{propagator-2}. While, in the case of IDT the function $f_{\rm IDT}(p)$ does not have any poles, being an exponential of an entire function. The IDT propagator in Eq.\eqref{propag-stringy} has an essential singularity at infinity, while the propagator in Eq.\eqref{propagator} does not diverge at infinity, along any direction. Another net distinction is related to the fact that the IDT propagator diverges in the UV for time-like exchange\footnote{In infinite derivative field theories, this pathologic behavior of the bare propagator can be cured by dressing it through quantum loop corrections \cite{Talaganis:2016ovm,Buoninfante:2018mre}.}, while the new nonlocal propagator does not, being well behaved for both time-like (s-channel) and space-like (t-channel) separations.

Both propagators have an improved UV behavior due to the presence of the exponential which gives a suppression in the high energy limit. Furthermore, the IDT action in Eq.\eqref{IDT-action-1} does not exhibits the enlarged Galilean symmetry in Eq.\eqref{larger-symmetry}.


\subsection{Comparison with $p$-adic string}

The action of $p$-adic string is given by \cite{Freund:1987kt}
\begin{equation}
S_{p{\rm -adic}}=\frac{m_s^4}{g_s^2}\frac{p^2}{p-1}\int d^4x \left(-\frac{1}{2}\phi p^{-\Box/m_s^2}\phi+\frac{1}{p+1}\phi^{p+1}\right),\label{p-adic-action-1}
\end{equation}
where $\phi(x)$ is a scalar field and represents the open string tachyon, the scale $m_s$ is the string mass and the parameter $g_s$ is the open string coupling constant. By defining $M_s^2\equiv m^2_s/{\rm ln}p$ we can also write the action in Eq.\eqref{p-adic-action-1} as follows
\begin{equation}
S_{p{\rm -adic}}=\frac{M_s^4}{g_s^2}\frac{p^2({\rm ln}p)^2}{p-1}\int d^4x \left(-\frac{1}{2}\phi e^{-\Box/M_s^2}\phi+\frac{1}{p+1}\phi^{p+1}\right).\label{p-adic-action-2}
\end{equation}
The first derivation of the $p$-adic action assumed $p$ to be a prime number, but it was realized that it can be defined for any positive integer, and even the case $p=1$ can be treated in some way \cite{Gerasimov:2000zp}. 

At first sight, the nonlocal action in Eq.\eqref{kineti term} seems a particular case of the $p$-adic action in Eq.\eqref{p-adic-action-2}, indeed for $p=1$ we obtain the term $\phi^2.$ However, the $p$-adic string for $p=1$ has nothing to do with the action in Eq.\eqref{kineti term}: the $1$-adic action also has a potential term as shown in Ref.\cite{Gerasimov:2000zp}, while the action in Eq.\eqref{kineti term} is purely kinetic. Moreover, unlike the propagator in Eq.\eqref{propagator}, the $p$-adic propagator is given by
\begin{equation}
\Pi_{p{\rm -adic}}(p)=e^{-p^2/M_s^2},\label{p-adic-propag}
\end{equation}
which possesses no poles. As for the high energy regime, both propagators in Eqs.(\ref{propagator},\ref{p-adic-propag}) have a similar ameliorated UV behavior.


\section{A nonlocal gravity with infinite complex conjugate poles}\label{new-gravity}

In this Section we will construct a linearized gravitational action around the Minkowski background by using the non-polynomial form-factor introduced in the previous sections, in such a way that the graviton propagator has a similar structure as the one in Eq.\eqref{propagator}. We will also find the corresponding non-linear action up to quadratic order in the curvature invariants, around Minkowski background. 

Let us consider the following gravitational action, which is the most general quadratic in the curvature, torsion free and parity invariant~\cite{Krasnikov,Kuzmin,Biswas:2005qr,Modesto:2011kw,Biswas:2011ar,Biswas:2016etb}~\footnote{See Ref.\cite{delaCruz-Dombriz:2018aal} for a more general infinite derivative action which also contains torsion and generalizes Poincaré gravity. Here, we will not discuss this more general case.}:
\begin{equation}
S= \frac{1}{16\pi G}\int d^4x\sqrt{-g}\left\lbrace \mathcal{R}+\frac{1}{2}\left(\mathcal{R}\mathcal{F}_1(\Box)\mathcal{R}+\mathcal{R}_{\mu\nu}\mathcal{F}_2(\Box)\mathcal{R}^{\mu\nu}+\mathcal{R}_{\mu\nu\rho\sigma}\mathcal{F}_3(\Box)\mathcal{R}^{\mu\nu\rho\sigma}\right)\right\rbrace ,
\label{quad-action}
\end{equation}
where $\mathcal{F}_i(\Box)$ are three form factors which we can be uniquely determined around Minkowski background by fixing the form of the graviton propagator~\cite{Modesto:2011kw,Biswas:2011ar}. Since we are interested in the linearized regime we can always neglect the Riemann squared term $\mathcal{R}_{\mu\nu\rho\sigma}\mathcal{F}_3(\Box)\mathcal{R}^{\mu\nu\rho\sigma}$; indeed the following identity holds  
\begin{equation}
\mathcal{R}_{\mu\nu\rho\sigma}\Box^n\mathcal{R}^{\mu\nu\rho\sigma}=4\mathcal{R}_{\mu\nu}\Box^n\mathcal{R}^{\mu\nu}-\mathcal{R}\Box^n\mathcal{R}+\mathcal{O}(\mathcal{R}^3)+{\rm div},
\end{equation}
where $\mathcal{O}(\mathcal{R}^3)$ takes into account of terms of the order $\mathcal{O}(h^3)$ and {\rm div} includes total derivatives.
By linearizing around the Minkowski,
\begin{equation}
g_{\mu\nu}=\eta_{\mu\nu}+\kappa h_{\mu\nu}\,,\label{lin-metric}
\end{equation}
where $\kappa:=\sqrt{8\pi G}$ and  $h_{\mu\nu}$ is the linearized metric perturbation, we obtain the following action up to $\mathcal{O}(h_{\mu\nu}^2)$ \cite{Biswas:2011ar}:
\begin{equation}
\begin{array}{rl}
S^{(2)}=&\displaystyle \frac{1}{4}\int d^4x\left\lbrace \frac{1}{2}h_{\mu\nu}f(\Box)\Box h^{\mu\nu}-h_{\mu}^{\sigma}f(\Box)\partial_{\sigma}\partial_{\nu}h^{\mu\nu}+hg(\Box)\partial_{\mu}\partial_{\nu}h^{\mu\nu}\right.\\[2.5mm]
&\,\,\,\,\,\,\,\,\,\,\,\,\,\,\,\,\,\,\,\,\,\,\,\,\displaystyle\left.-\frac{1}{2}hg(\Box)\Box h+\frac{1}{2}h^{\lambda\sigma}\frac{f(\Box)-g(\Box)}{\Box}\partial_{\lambda}\partial_{\sigma}\partial_{\mu}\partial_{\nu}h^{\mu\nu}\right\rbrace,
\label{lin-quad-action}
\end{array}
\end{equation}
where $h\equiv\eta_{\mu\nu}h^{\mu\nu}$ is the trace, $\Box=\eta^{\mu\nu}\partial_{\mu}\partial_{\nu}$ is the flat d'Alembertian and
\begin{equation}
f(\Box)=\displaystyle  1+\frac{1}{2}\mathcal{F}_2(\Box)\Box,\,\,\,\,\,\,\,\,\,\,\,\,\,
g(\Box)= 1-2\mathcal{F}_1(\Box)\Box-\frac{1}{2}\mathcal{F}_2(\Box)\Box\,.
\end{equation}
The gauge independent part of the saturated graviton propagator around the Minkowski background has the following general expression \cite{Krasnikov,Modesto:2011kw,Biswas:2011ar}:
\begin{equation}
\Pi(p)=\frac{\mathcal{P}^2}{f(p)p^2}+\frac{\mathcal{P}_s^0}{(f(p)-3g(p))p^2},
\end{equation}
where we have suppressed the tensorial indices for simplicity; $\mathcal{P}^2$ and $\mathcal{P}^0_s$ are the so called spin projection operators, which project along the spin-$2$ and spin-$0$ components, respectively; see Refs.\cite{Biswas:2011ar,Biswas:2013kla,Buoninfante} for more details. Note that for $f=1=g$ we recover the graviton propagator, $\Pi_{GR}(p)=\mathcal{P}^2/p^2-\mathcal{P}^0_s/2p^2,$ of the Einstein general relativity.


\subsection{Gravitational action and propagator}

The gravitational analog of the nonlocal scalar propagator defined in Eq.\eqref{propagator} can be found by making the following choice for the functions $f(p)$ and $g(p):$
\begin{equation}
f(p)=g(p)=M_s^2\frac{e^{p^2/M_s^2}-1}{p^2}\,\,\longrightarrow\,\,f(\Box)=g(\Box)=-M_s^2\frac{e^{-\Box/M_s^2}-1}{\Box},\label{choice}
\end{equation}
which fixes the form of the gauge independent part of the saturated propagator as follows
\begin{equation}
\Pi(p)=\frac{1}{f(p)}\Pi_{GR}(p)=\frac{p^2}{M_s^2\left(e^{p^2/M_s^2}-1\right)}\Pi_{GR}(p)=\frac{1}{M_s^2\left(e^{p^2/M_s^2}-1\right)}\left(\mathcal{P}^2-\frac{\mathcal{P}^0_s}{2}\right).\label{nonlocal-grav-propag}
\end{equation}
Note that the linearized action in Eq.\eqref{lin-quad-action} with the nonlocal operator in Eq.\eqref{choice} is invariant under the following enlarged Galilean transformation\footnote{Note that the transformation in Eq.\eqref{enlarged-galil-gravity} is uniquely defined up to constant factors. Indeed, the transformations $\delta h_{\mu\nu}=c_{\mu\nu}\psi_k,$ with $c_{\mu\nu}$ being a constant symmetric tensor, and $\delta h_{\mu\nu}=c\partial_{\mu}\partial_{\nu}\psi_k$ with $c$ being a constant scalar, have the same functional form as the one in Eq.\eqref{enlarged-galil-gravity}. It is obvious that this enlarged Galilean symmetry is only respected at the linearized level, while it is not realized at the non-linear level.}:
\begin{equation}
h_{\mu\nu}(x)\longrightarrow h_{\mu\nu}(x)+\eta_{\mu\nu}\psi_k(x),\label{enlarged-galil-gravity}
\end{equation}
where $\psi_k$ has been introduced in Eqs.(\ref{boundary-cond-4D},\ref{compact-form}).

Also in this case, the propagator possesses a massless real pole, $p^2=0,$ and an infinite number of massive complex conjugate poles. However, as we have shown in Subsection \ref{unitarity-subsection} tree-level unitarity is maintained. From the choice made in Eq.\eqref{choice}, we can also find the expressions for the form factors in the full non-linear gravitational action, indeed we can easily check that Eq.\eqref{choice} implies
\begin{equation}
2\mathcal{F}_1(\Box)=-\mathcal{F}_2(\Box)=2M_s^2\frac{e^{-\Box/M_s^2}-1}{\Box^2}+\frac{2}{\Box}.\label{form-factors}
\end{equation}
Thus, the nonlocal quadratic gravitational action is given by the following expression:
\begin{equation}
S= \frac{1}{16\pi G}\int d^4x\sqrt{-g}\left\lbrace \mathcal{R}- G_{\mu\nu}\frac{1}{\Box}\mathcal{R}^{\mu\nu}- M_s^2G_{\mu\nu}\frac{e^{-\Box/M_s^2}-1}{\Box^2}\mathcal{R}^{\mu\nu}\right\rbrace ,
\label{nonlocal-quad-action}
\end{equation}
where $G_{\mu\nu}=\mathcal{R}_{\mu\nu}-1/2g_{\mu\nu}\mathcal{R}$ is the Einstein tensor. Note that the two quadratic terms with $1/\Box$ and $1/\Box^2$ are individually non-analytic, but their particular combination in Eq.\eqref{form-factors}, and so the gravitational action in Eq.\eqref{nonlocal-quad-action}, are analytic functions of $\Box.$ Such an analiticity is also respected at the linearized level, where the nonlocal operator in the kinetic term is given by $\left(e^{-\Box/M_s^2}-1\right)/\Box;$ see Eq.\eqref{lin-field-eq} of the next subsection. In relation with the discussion in Subsection \ref{unitarity-subsection}, the nonlocal gravitational model in Eq.\eqref{nonlocal-quad-action} can be seen as an infinite derivative generalization of (local) Lee-Wick theories of gravity \cite{Anselmi:2017ygm,Modesto:2015ozb,Accioly:2016qeb}.


\subsection{Linearized metric solution}

We now wish to compute the gravitational potential generated by a point-like source in the nonlocal theory of gravity introduced above. By varying the linearized action in Eq.\eqref{lin-quad-action}, and imposing the choice in Eq.\eqref{choice}, we can obtain the following linearized field equations:
\begin{equation}
\displaystyle M^2_s\frac{e^{-\Box/M_s^2}-1}{\Box}\left(\Box h_{\mu\nu}-\partial_{\sigma}\partial_{\nu}h_{\mu}^{\sigma}-\partial_{\sigma}\partial_{\mu}h_{\nu}^{\sigma}+\eta_{\mu\nu}\partial_{\rho}\partial_{\sigma}h^{\rho\sigma}
+\partial_{\mu}\partial_{\nu}h-\eta_{\mu\nu}\Box h\right)=16\pi G T_{\mu\nu}\,,
\label{lin-field-eq}
\end{equation}
where $T_{\mu\nu}$ is the stress-energy tensor of the matter sector. 


By working in the transverse gauge, we can write a static and spherically symmetric linearized metric as follows \cite{Carroll:2004st}
\begin{equation}
ds^2=-(1+2\Phi)dt^2+(1-2\Psi)(dr^2+r^2d\Omega^2),\label{isotr-metric}
\end{equation}
so that  $\kappa h_{00}=-2\Phi<1$, $\kappa h_{ij}=-2\Psi\delta_{ij}<1$, $\kappa h=2(\Phi-3\Psi),$ with $\Phi$ and $\Psi$ being the two unknown metric potentials. Then, by considering a static point-like source, of mass $m,$ described by the stress-energy tensor $T_{\mu\nu}=m\delta_{\mu}^0\delta_{\nu}^{0}\delta^{(3)}(\vec{r}),$ one can show that the metric potentials solve the following modified Poisson equation:
\begin{equation}
M_s^2\left(e^{-\nabla^2/M_s^2}-1\right)\Phi(r)=M_s^2\left(e^{-\nabla^2/M_s^2}-1\right)\Psi(r)=-4\pi Gm\delta^{(3)}(\vec{r}),
\label{field-eq-pot}
\end{equation}
which in the local limit $M_s\rightarrow \infty$ recovers the standard Poisson equation for the Newtonian potential.
The solution of the nonlocal differential equation in Eq.\eqref{field-eq-pot} can be found by going to Fourier space and then anti-transforming back to coordinate space as follows
\begin{equation}
\Phi(r)=\Psi(r)= \displaystyle -\frac{2 Gm}{\pi M_s^2}\frac{1}{r}\int_0^{\infty}dk\frac{k\,{\rm sin}(kr)}{e^{k^2/M_s^2}-1}\,.
\label{fourier-pot}
\end{equation}
We have not been able to compute the integral in Eq.\eqref{fourier-pot} analytically, however we can do it numerically and its behavior is shown in comparison with the Newtonian potential\footnote{It is worthwhile mentioning that the best experiment of Newton's law has been performed with torsion balances and the law has been tested up to $5.6\times 10^{-5}$m \cite{-D.-J.}. From such an experiment, one could infer a rough constraint on the new scale which would read $1/M_s< 5.6\times 10^{-5}$m.} in Fig. $2.$ 

\begin{figure}[t]
	\includegraphics[scale=0.50]{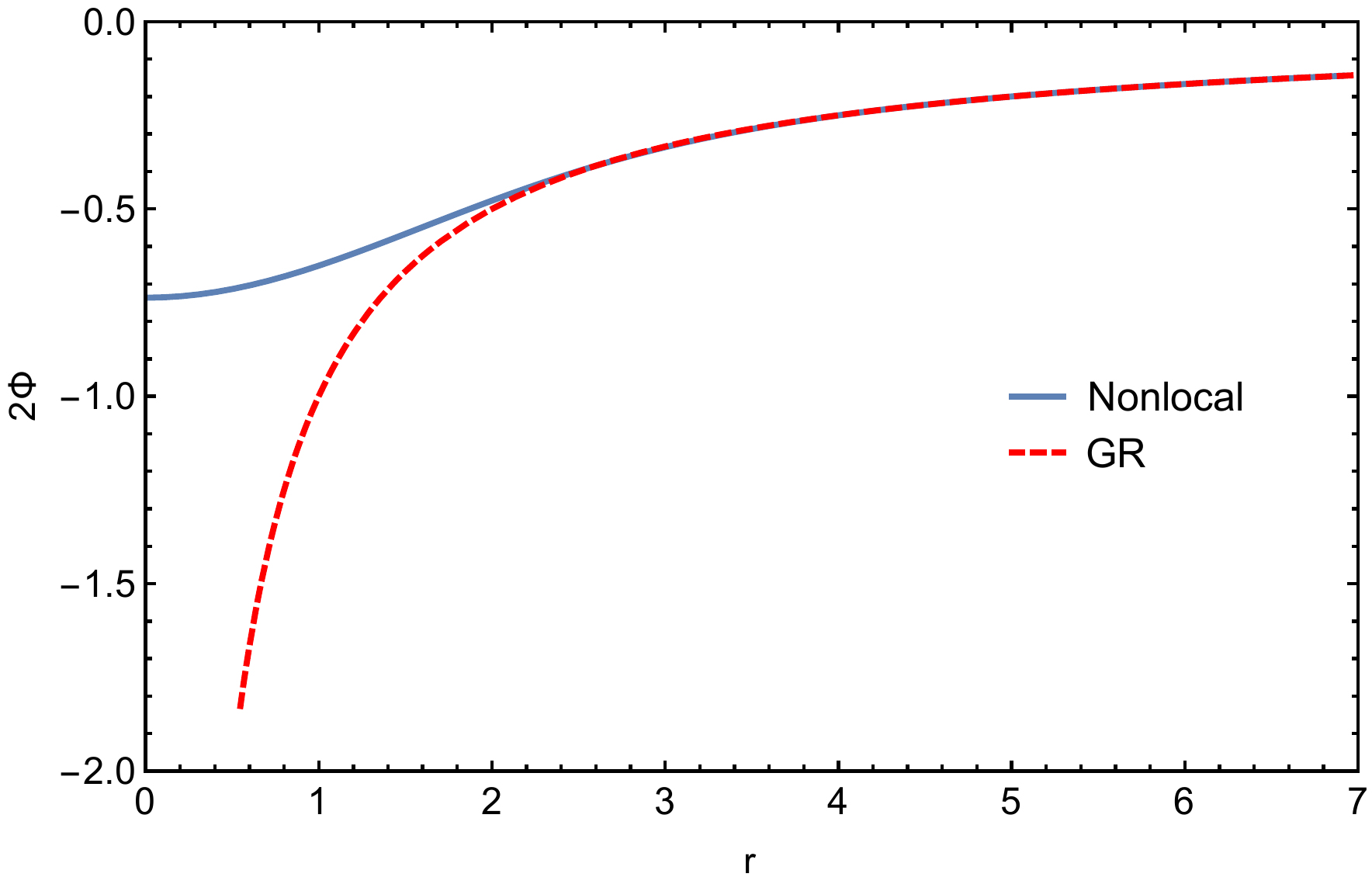}
	\centering
	\protect\caption{We have plotted the numerical result of the integral in Eq.\eqref{fourier-pot} (blue solid line), which represents the behavior of the gravitational metric potential generated by a point-like source in the nonlocal theory described by the action in Eq.\eqref{nonlocal-quad-action}. We have also plotted the Newtonian potential (red dashed line) to make a comparison. We have set $G=1=M_s$ and $m=0.5$ for simplicity.}
\end{figure}

We can notice that the metric potential is nonsingular at the origin, unlike the Newtonian one, indeed nonlocality regularizes its behavior. We can exactly compute the value of the metric potential at $r=0,$ and is given by:
\begin{equation}
\Phi(0)\equiv \lim_{r\rightarrow0}\Phi(r)=-\frac{2 Gm}{\pi M_s^2}\int\limits_{0}^{\infty} dk \frac{k^2}{e^{k^2/M_s^2}-1}=-\frac{GmM_s}{2\sqrt{\pi}}\zeta\left(\frac{3}{2}\right),
\label{fourier-pot-r=0}
\end{equation}
where $\zeta(x)\equiv \sum\limits_{n=0}^{\infty}n^{-x}$ is the {\it Riemann zeta function}. Note that the linearity of the metric potential holds as long as the following inequality is satisfied:
\begin{equation}
\frac{GmM_s}{\sqrt{\pi}}\zeta\left(\frac{3}{2}\right)< 1\,.
\end{equation}


\subsection{Comparison with infinite derivative gravity (IDG)}

Now we briefly make a comparison between the nonlocal gravity model we have introduced above and the infinite derivative gravity (IDG) studied in Refs.\cite{Biswas:2005qr,Modesto:2011kw,Biswas:2011ar}. The simplest form of the gravitational action in IDG is given by the following choice for the form factors \cite{Modesto:2011kw,Biswas:2011ar,Biswas:2013cha}:
\begin{equation}
f_{\rm IDG}(\Box)=g_{\rm IDG}(\Box)=e^{-\Box/M_s^2}\, \Longrightarrow\, 2\mathcal{F}_{{\rm IDG},1}(\Box)=-\mathcal{F}_{{\rm IDG},2}(\Box)=-2\frac{e^{-\Box/M_s^2}-1}{\Box},\label{form-factors-IDG}
\end{equation}
which fixes the form of the gravitational action as follows
\begin{equation}
S_{\rm IDG}= \frac{1}{16\pi G}\int d^4x\sqrt{-g}\left\lbrace \mathcal{R}+ G_{\mu\nu}\frac{e^{-\Box/M_s^2}-1}{\Box}\mathcal{R}^{\mu\nu}\right\rbrace .
\label{nonlocal-quad-action-IDG}
\end{equation}
The gauge independent part of the graviton propagator around the Minkowski background takes the following form \cite{Krasnikov,Tomboulis:1997gg,Modesto:2011kw,Biswas:2005qr}:
\begin{equation}
\Pi_{\rm IDG}(p)=\frac{1}{f_{\rm IDG}(p)}\Pi_{GR}(p)=e^{-p^2/M_s^2}\Pi_{GR}(p),\,\,\,\,\,\,\,\,\,\,f_{\rm IDG}(p)\equiv e^{p^2/M_s^2}.\label{nonlocal-grav-propag-IDG}
\end{equation}
As in the scalar case analyzed in Section \ref{enlarging}, we note that there is a net difference between the two propagators in Eqs.(\ref{nonlocal-grav-propag},~\ref{nonlocal-grav-propag-IDG}). The main difference is contained in the function $f(p).$ Indeed, in IDG the function $f_{\rm IDG}(p)$ does not introduce any new degrees of freedom besides the massless spin-$2$ graviton. While, for the gravitational model proposed in this paper, the function $f(p)$ possesses zeros which introduce extra poles in the graviton propagator in Eq.\eqref{nonlocal-grav-propag}, and they are complex and infinite in number. Nevertheless, both satisfy the tree-level unitarity. 

\section{Conclusions}\label{conclus}

In this paper we have introduced a new nonlocal model motivated by a symmetry principle. We have noticed that by working with nonlocal actions it is possible to enlarge the Galilean
shift symmetry. Indeed, we have found a new class of transformations,
$\phi\longrightarrow \phi +\psi_k$ (see Eq.\eqref{larger-symmetry}),
under which our nonlocal actions are invariant, and such a wider
class includes the Galilean shift symmetry as a subclass, for $k=0.$  The energy scale $M_s$ does not play the role of a cut-off but it is a new physical scale which tells us when nonlocality becomes relevant. In this sense nonlocal theories are an attempt towards UV completeness and are aimed to go beyond the effective field theory prescription.

For a scalar field we have constructed some possible actions which
respect such a symmetry, and studied the corresponding field equation
and propagator. The pole structure of the propagator turns out to be
more complicated; indeed, besides the massless pole of the local
theory, we have also an infinite number of complex conjugate poles
which, in principle, may spoil unitarity. However, we have noticed
that all the infinite complex poles do not contribute to the imaginary
part of the propagator, so that unitarity turns out to be preserved at
the tree-level. We have good hints that the optical theorem holds at
all orders in perturbation theory, but further investigations will be
subject of future works.  Moreover, we have made some comparisons with
other two already well known nonlocal theories, i.e. infinite
derivative field theory and $p$-adic string. 
Our new nonlocal model can be seen as an infinite derivative
generalization of Lee-Wick theories, as we have propagators with pairs
of conjugate poles which are not finite in number, but infinite.

We have analyzed these particular form factors also in the context
of linearized gravity. We have constructed its nonlocal action
quadratic in the curvature, which in the linear regime around Minkowski
manifestly exhibits the enlarged Galilean symmetry. Also in this
case we have computed the propagator which, besides the massless
graviton pole, possesses an infinite number of complex
conjugate poles, but no real ghost-modes. We have also computed the
gravitational potential generated by a point-like source, which
turns out to be nonsingular at the origin, unlike the Newtonian
one. 

In the framework of nonlocal theories, the requirement of using exponentials of entire functions seemed to be fundamental in order to avoid ghost-modes in higher derivative gravity\footnote{It is worthwhile to mention that the authors in Refs.\cite{Anselmi:2018ibi} introduced a new quantization prescription which converts the ghost-mode into a so called {\it fakeon} (fake particle), in such a way that both unitarity and renormalizability are satisfied even in $4$-derivative gravity.} \cite{Biswas:2005qr,Biswas:2011ar} and, thus, preserving unitarity at the perturbative level \cite{sen-epsilon,carone,Briscese:2018oyx,chin}. Very interestingly, we have been able to enlarge the
class of nonlocal form factors which make the propagator
ghost-free, without restricting ourself to functions $f(p)$ which
are equal to exponential of entire functions. However, further studies to show that the optical theorem holds at each order in perturbation theory will be subject of future works.

Analogously to the
scalar field case, this new model of nonlocal gravity can be seen as
an infinite derivative generalization of Lee-Wick theories of
gravity. However, we have to be careful in the extension to full
non-linear gravity because Galilean shift symmetry must be broken
when we couple it to gravity even in finite order derivative
theories. We need to consider further extension
\cite{Pirtskhalava:2015nla,Gabadadze:2012tr}, which will be investigated in future publication. These new form factors might be useful to be explored also in the context of string field theory.

Finally, one could also take a different point of view according to which nonlocal extensions of the Galilean symmetry might be useful to formulate nonlocal effective field theories, analogously to the local case with the Galilean symmetry, but now by demanding the enlarged symmetry in Eq.\eqref{larger-symmetry} to be satisfied. This might be crucial for discriminating local and nonlocal theories. We leave such a possibility for future investigations.




\acknowledgements
The authors thank Francesco Di Filippo, Anupam Mazumdar and Luciano Petruzziello for constructive comments. M.Y. is supported in part by JSPS KAKENHI Grant Numbers JP25287054, JP15H05888, JP18H04579, JP18K18764, and by the Mitsubishi Foundation. M.Y. would like to thank Luca Buoninfante and Gaetano Lambiase for invitation and hospitality during his stay at University of Salerno, where this work was completed.


\appendix

\section{Enlarging the shift symmetry}\label{append-shift}

In this paper we have constructed nonlocal Lagrangians invariant under the transformations in Eq.\eqref{larger-symmetry}, which generalize the Galilean symmetry in Eq.\eqref{galilean-transf-2} to the case of non-polynomial differential operators. However, we have not yet considered the simpler case of the {\it shift symmetry}, which is defined by the following transformation \cite{Freese:1990rb,Kawasaki:2000yn,Kawasaki:2000ws}:
\begin{equation}
\phi(x)\longrightarrow \phi(x)+c,
\label{shift-transf}
\end{equation}
where $c$ is a constant parameter generating the whole family of shift transformations; for instance, first derivative operators like $\partial_{x}\phi$ exhibit such a symmetry. As we have done for the Galilean symmetry in Section \ref{enlarging}, we now wish to find some differential operator invariant under a wider class which also includes the shift symmetry in Eq.\eqref{shift-transf} as a subclass. In other words, we want to find the form of the function $\chi(x)$ such that the operators are invariant under the transformation $\phi\longrightarrow\phi+\chi,$ but $\partial_x\chi\neq0.$

Let us consider a $1$-dimensional case and find the solution for the following differential equation
\begin{equation}
\left(e^{-\partial_x/M_s}-1\right)\chi(x)=0 \Longleftrightarrow e^{-\partial_x/M_s}\chi(x)=\chi(x), \label{shift nonlocal-1D}
\end{equation}
which also means to find the eigenfunctions $\chi$ of the operator $e^{-\partial_x/M_s}$ of unit eigenvalue.
First of all, note that the following property holds 
\begin{equation}
e^{-\partial_x/M_s}=e^{-\partial_x/M_s+i2\pi k}\equiv e^{\theta_{1}}, \label{property-shift}
\end{equation}
where $k$ is an integer number and we have defined the differential operator
\begin{equation}
\theta_{1}:=-\frac{\partial_x}{M_s}+i2\pi k. \label{1-diff-oper-shift}
\end{equation}
Thus, we need to find solutions for the equation $\theta_{1}\chi(x)=0:$
\begin{equation}
\theta_{1} \chi(x)=-\frac{\partial_x}{M_s}\chi(x)+i2\pi k\chi(x)=0\,\Longrightarrow \,\chi_k(x)=c e^{i\pi k M_sx}. \label{1-sol-shift}
\end{equation}
where $c$ is an integration constant, which can be uniquely fixed by imposing that for $k=0$ we recover the shift symmetry in Eq.\eqref{shift-transf}.

Hence, we have explicitly shown that by working with the non-polynomial operator introduced in Eq.\eqref{shift nonlocal-1D}, we can enlarge the shift symmetry, which now becomes a subclass of a wider class described by the following family of parameters: 
\begin{equation}
\left\lbrace c,k\right\rbrace,\label{family-parameters-shift}
\end{equation}
which is expressed in terms of the following field transformation:
\begin{equation}
\phi(x)\longrightarrow \phi(x)+\chi_k(x).\label{larger-shift}
\end{equation}
Thus, the shift transformations correspond to the subfamily $\left\lbrace c,0\right\rbrace.$ 

The generalization to a $4$-dimensional case is given by replacing the operator $\partial_x$ by $d^{\mu} \partial_{\mu}$ with $d_{\mu}$ being a constant vector: such a non-polynomial operator might break the Lorentz invariance explicitly.

\subsection{A nonlocal generalization of the Dirac action}

Another interesting case with this enlarged shift symmetry in $4$ dimension is a fermion, $\psi(x)$. We want to find the form of the function $\lambda(x)$ such that the operators are invariant under the transformation $\psi(x)\longrightarrow\psi(x)+\lambda(x),$ but $\partial_{\mu}\lambda(x)\neq 0.$

Let us find the solution for the following differential equation
\begin{equation}
\left(e^{i\gamma^{\mu}\partial_{\mu}/M_s}-1\right)\lambda(x)=0 \Longleftrightarrow e^{i\gamma^{\mu}\partial_{\mu}/M_s}\lambda(x)=\lambda(x), \label{shift nonlocal-fermion}
\end{equation}
which means to find the eigenfunctions $\psi(x)$ of the operator $e^{i\gamma^{\mu}\partial_{\mu}/M_s}$ of unit eigenvalue; here, $\gamma^{\mu}$ are the standard Gamma matrices. Note that the following property holds 
\begin{equation}
e^{i\gamma^{\mu}\partial_{\mu}/M_s}=e^{i\gamma^{\mu}\partial_{\mu}/M_s+i2\pi k \mathbb{1}}\equiv e^{\hat{\theta}_{1}}, \label{property-shift-fermion}
\end{equation}
where $k$ is an integer number, $\mathbb{1}$ is the $4\times4$ identity matrix, and we have defined the differential matrix operator
\begin{equation}
\hat{\theta}_{1}:=i\frac{\gamma^{\mu}\partial_{\mu}}{M_s}+i2\pi k \mathbb{1}. \label{1-diff-oper-shift-fermion}
\end{equation}
Thus, we need to find solutions for the equation $\hat{\theta}_{1}\lambda(x)=0:$
\begin{equation}
\hat{\theta}_{1} \lambda(x)=i\frac{\gamma^{\mu}\partial_{\mu}}{M_s}\lambda(x)+i2\pi k \lambda(x)=0\,\Longrightarrow \,\lambda_k(x)= e^{i M_s l_{\mu}x^{\mu}} \lambda_0. \label{1-sol-shift-fermion}
\end{equation}
where $\lambda_0$ is a constant Dirac spinor, which can be uniquely fixed by imposing that for $k=0$ we recover the standard shift symmetry, while $l_{\mu}$ is a vector satisfying $-l_{\mu} \gamma^{\mu} / M_s + 2\pi i k \mathbb{1} = 0$.

Hence, we have explicitly shown that by working with the non-polynomial operator introduced in Eq.\eqref{shift nonlocal-fermion}, we can enlarge the shift symmetry, which now becomes a subclass of a wider class described by the following family of parameters: 
\begin{equation}
\left\lbrace \lambda_0,k\right\rbrace,\label{family-parameters-shift-fermion}
\end{equation}
which is expressed in terms of the following field transformation:
\begin{equation}
\psi(x)\longrightarrow \psi(x)+\lambda_k(x).\label{larger-shift-fermion}
\end{equation}
Thus, the shift transformations correspond to the subfamily $\left\lbrace \lambda_0,0\right\rbrace.$ 

Note that the following action is invariant under this enlarged shift symmetry:
\begin{equation}
S = M_s\int d^4x  \bar{\psi} \left(e^{i\gamma^{\mu}\partial_{\mu}/M_s}-1\right) \psi,\label{nonlocal-dirac}
\end{equation}
which is a natural nonlocal extension of the Dirac action. Such an action is invariant under the global $U(1).$ Furthermore, if we introduce a gauge field $A_{\mu}$ and covariantize the derivatives, the following operator $e^{i\gamma^{\mu}D_{\mu}},$ with $D_{\mu}=\partial_{\mu}+ieA_{\mu},$ one can show that also the local gauge symmetry is preserved. Indeed, if $\psi\longrightarrow e^{i\alpha(x)}\psi$ and $A_{\mu}\longrightarrow A_{\mu}+1/e\partial_{\mu}\alpha(x),$ then $\left(i\gamma^{\mu}D_{\mu}\right)^n\psi\longrightarrow e^{i\alpha(x)}\left(i\gamma^{\mu}D_{\mu}\right)^n\psi.$

The fermionic nonlocal action in Eq.\eqref{nonlocal-dirac} will be further investigated in future publication.


\end{document}